\documentstyle[11pt,aaspp4]{article}

\slugcomment{To appear in the Astronomical Journal (accepted 1998 August 10)}

\lefthead{Mighell, Sarajedini, \& French}
\righthead{Oldest Star Clusters in the SMC}

\begin{document}

\def\et{\hbox{\em{et~al.\ }}}
\def\vs{\hbox{vs }}
\def\feh{\hbox{[Fe/H]}}
\def\bmv{\hbox{$B\!-\!V$}}
\def\ebmv{\hbox{E$(\bmv)$}}
\def\dbmv{\hbox{$d_{(B\!-\!V)}$}}
\def\dbmr{\hbox{$d_{(B\!-\!R)}$}}
\def\mvto{\hbox{$M_V^{\rm{TO}}$}}
\def\mvrhb{\hbox{$M_V^{\rm{RHB}}$}}
\def\vrhb{\hbox{$V_{\rm{RHB}}$}}
\def\vhb{\hbox{$V_{\rm{HB}}$}}
\def\lea{\mathrel{<\kern-1.0em\lower0.9ex\hbox{$\sim$}}}
\def\gea{\mathrel{>\kern-1.0em\lower0.9ex\hbox{$\sim$}}}
\def\ifundefined#1{\expandafter\ifx\csname#1\endcsname\relax}
\def\firstuse#1{\marginpar[]{{\scriptsize\hspace*{-1.5mm}$\bullet${\sc{#1}}}}}

\title{WFPC2 OBSERVATIONS OF STAR CLUSTERS IN THE MAGELLANIC CLOUDS.
II. THE OLDEST STAR CLUSTERS IN THE SMALL MAGELLANIC CLOUD\altaffilmark{1}
}

\author{
\sc
Kenneth J. Mighell\altaffilmark{2}
}
\affil{
\small
Kitt Peak National Observatory,
National Optical Astronomy Observatories\altaffilmark{3},\\
P. O. Box 26732, Tucson, AZ~~85726-6732\\
Electronic mail:
mighell@noao.edu
}

\author{
\sc
Ata Sarajedini\altaffilmark{4}
}
\affil{
\small
Department of Physics and Astronomy,
San Francisco State University,
1600 Holloway Avenue,
San Francisco, CA~~94132\\
Electronic mail:
ata@stars.sfsu.edu
}

\author{
{\sc
Rica S. French\altaffilmark{5}
}
}
\affil{Middle Tennessee State University,
Physics \& Astronomy Department,
WPS 219, P. O. Box 71, Murfreesboro, TN~~37132\\
Electronic mail:
sirbaugh@physics.mtsu.edu
}

\altaffiltext{1}{
Based on observations made with the NASA/ESA
{\em{Hubble Space Telescope}},
obtained from the data archive at the Space Telescope
Science Institute,
which is operated by the Association of
Universities for Research in Astronomy, Inc.\ under NASA
contract NAS5-26555.
}

\altaffiltext{2}{
Guest User, Canadian Astronomy Data Centre, which is operated by the
Dominion Astrophysical Observatory for the National Research Council of
Canada's Herzberg Institute of Astrophysics.
}

\altaffiltext{3}{NOAO is operated
by the Association of Universities for Research in Astronomy, Inc., under
cooperative agreement with the National Science Foundation.}

\altaffiltext{4}{Hubble Fellow}

\altaffiltext{5}{Based on research conducted at NOAO as part of the
Research Experiences for Undergraduates program.}

\newpage
\begin{abstract}

We present our analysis of archival
{\sl{Hubble Space Telescope}} Wide Field Planetary Camera 2 (WFPC2)
observations
in
F450W ($\sim$$B$)
and
F555W ($\sim$$V$)
of the intermediate-age populous star clusters
NGC 121, NGC 339, NGC 361, NGC 416,
and Kron 3 in the
Small Magellanic Cloud.
We use published photometry of two other SMC populous star clusters,
Lindsay 1 and Lindsay 113, to investigate
the age sequence of these seven populous star clusters
in order to improve our understanding of the
formation chronology of the SMC.
We analyzed the
$V$ vs $\bmv$ and $M_V$ vs $(\bmv)_o$ color-magnitude diagrams
of these populous Small Magellanic Cloud star clusters
using a variety of techniques and
determined their
ages,
metallicities,
and
reddenings.
These new data enable us to improve
the age-metallicity relation of star clusters
in the Small Magellanic Cloud.
In particular,
we find that a closed-box continuous star-formation model
does not reproduce the age-metallicity relation adequately.
However,
a theoretical model punctuated by bursts of star formation
is in better agreement with the observational data presented herein.
\end{abstract}


\section{INTRODUCTION}

We can improve our understanding of
the early chemical enrichment history of the Small Magellanic Cloud
(SMC) by investigating the ages and abundances of its oldest star clusters.
A recent summary of the SMC age-metallicity relation may be found in
Fig.~1a of the review article by
Olszewski, Suntzeff, \& Mateo (\cite{olet1996}).
In an earlier review article, Da Costa (\cite{daco1991}) notes
that the most metal-poor clusters in the SMC have similar
metallicities yet apparently very different ages indicating that this
galaxy has had an
unusual chemical enrichment history.
The major caveat to this conclusion was that the observational errors were
large; typical error bars in $\feh$
were between $\pm$0.2 and $\pm$0.4 dex while the age errors were between
$\pm$2 and $\pm$3 Gyr.  This has lead to significant uncertainty in
our present understanding of the age-metallicity relationship
for the SMC.

It has only recently been established that the
most metal-poor globular clusters in the Large Magellanic Cloud (LMC)
have ages that are comparable to the ages of the metal-poor Galactic
globular clusters
(Hodge 11:
Mighell \et \cite{miet1996};
NGC
1786,
1841,
2210:
Brocato \et \cite{bret1996};
NGC 1754,
1835,
1898,
2005,
2109:
Olsen \et \cite{olet1998}).
The populous LMC star cluster
ESO 121-SC03 has an age of $\sim$$9$ Gyr
and is the only intermediate-age star cluster in the LMC within the
age range of 3 and 12 Gyr
(Geisler \et \cite{geet1997}, and references therein).
Recent work suggests that ESO 121-SC03 may not be alone
(Sarajedini \cite{sa1998}).

The SMC has at least 7 populous star
clusters that are metal-poor ($\feh \lea -1.0$)
with ages between $\sim$$5$ and $\sim$$12$ Gyr:
Lindsay 113, Kron 3, NGC 339, NGC 416, NGC 361, Lindsay 1, NGC 121.
All these populous clusters have similar metallicities
and horizontal branch (HB) morphologies that are
predominantly redward of the RR Lyrae instability strip.
In our recent investigation of three SMC intermediate-age populous
clusters (NGC 416, Lindsay 1, Lindsay 113)
we showed how existing SMC cluster ages in the literature
derived from integrated photometry can be very unreliable
(Mighell, Sarajedini, \& French \cite{miet1998}; hereafter {\sc{PaperI}}).
Geisler \et (\cite{geet1997}) found similar problems with
published LMC cluster ages derived from integrated photometry.
The unintended inclusion of bright field main-sequence stars
and the poor age sensitivity provided by broad-band colors
for star clusters older than a few Gyr are the two principle reasons
why integrated cluster photometry can confuse intermediate-age
clusters with genuine old clusters.

The more massive LMC has received considerably more
attention by stellar population researchers than has the
SMC.
The early photographic studies of NGC 361 (Arp \cite{ar1958})
and NGC 339 (Gascoigne \cite{ga1966})
have the only color-magnitude diagrams of these SMC clusters
available in the refereed literature.
Most of the clusters that do have CCD-based color-magnitude diagrams are
from studies that used early CCD cameras in the 1980's.
Modern cluster observations
using large-format low-noise CCDs can significantly improve our
knowledge of the oldest star clusters in the Small Magellanic Cloud.
The analysis of color-magnitude diagrams produced from such studies
can provide cluster age and metallicity data which
complements and enhances chemical abundance studies based on
spectroscopic observations of cluster red giants
(e.g., Da Costa \& Hatzidimitriou \cite{daha1998}, hereafter DH98).

In this work, we improve our understanding of
the early chemical enrichment history of the Small Magellanic Cloud
using archival WFPC2 data and previously published photometry of
its oldest star clusters.
Section 2 is a discussion of the observations and photometric reductions.
We present our color-magnitude diagrams of the SMC star
clusters NGC 416, NGC 121, NGC 339, NGC 361, and Kron 3 in Sec.\ 3.
We estimate the reddenings and metallicites of the oldest SMC star clusters
in Sec.\ 4 and estimate their ages in Sec.\ 5.
Section 6 is a discussion of the early star formation history of the
SMC galaxy.
We then present our conclusions and thoughts about future studies
of the stellar populations of the Small Magellanic Cloud
in the final section of this paper.

\section{OBSERVATIONS AND PHOTOMETRIC REDUCTIONS}

The SMC populous star clusters
NGC 121, NGC 339, NGC 361, NGC 416, and Kron 3
were observed with the
{\sl{Hubble Space Telescope}} Wide Field Planetary Camera 2 (WFPC2)
between 1994 January 26 and 1994 May 27 through the
F450W ($\sim$$B$)
and F555W ($\sim$$V$)
filters.
The WFPC2 PC1 aperture
(Biretta \et \cite{biet1996})
was centered on the target positions
given in
Table\ \ref{tbl-obslog}\firstuse{Tab\ref{tbl-obslog}}
and shown in
Fig.\ \ref{fig-1}\firstuse{Fig\ref{fig-1}}.
Ten high-gain observations were obtained.
These WFPC2 datasets were recalibrated at the Canadian Astronomy Data Centre
and retrieved electronically by us using a guest account which was
kindly established for KJM.

The data were analyzed with
the CCDCAP\footnote{
IRAF implementations of CCDCAP are now available over the Wide World Web
at the following site:~
{\tt{http://www.noao.edu/noao/staff/mighell/ccdcap/}}
}
digital circular aperture
photometry code developed by Mighell
to analyze
{\sl{HST}} WFPC2
observations.
We followed the same WFPC2 data reduction procedures
of Mighell (\cite{mi1997}) except for the following details.
A fixed aperture with a radius of 2.0 pixels was used for all stars
on the WFPC2 CCDs.
The local background level was determined from a robust estimate
of the mean intensity value of all pixels between 2.0 and 5.5 pixels
from the center of the circular stellar aperture.
The Charge Transfer Effect was
removed from the instrumental magnitudes
by using a 4\% uniform wedge
along the Y-axis of each CCD as described in
Holtzman \et (\cite{hoet1995b}).
We used the standard WFPC2 magnitude system
(Holtzman \et \cite{hoet1995b})
which is defined using 1\arcsec\ diameter apertures
containing about 90\% of the total flux from a star.
The instrumental magnitudes,
$b_r$ and $v_r$,
were transformed to Johnson $B$ and $V$ magnitudes
using the following equations
$
B
=
b_r
+ \Delta_r
+ \delta_r
+ [ 0.230\!\pm\!0.006](\bmv)
+ [-0.003\!\pm\!0.006](\bmv)^2
+ [21.175\!\pm\!0.002]
$
and
$
V
=
v_r
+ \Delta_r
+ \delta_r
+ [-0.060\!\pm\!0.006](\bmv)
+ [ 0.033\!\pm\!0.002](\bmv)^2
+ [21.725\!\pm\!0.004]
$
where an instrumental magnitude of zero is defined as one DN/sec at
the high gain state ($\sim$14 e$^-$/DN).
The constants come from
Tables 10 and 7, respectively, of Holtzman \et (\cite{hoet1995b}).
These color equations were used by
Mighell \et (\cite{miet1996})
to show that the age of Large Magellanic Cloud cluster
Hodge 11 is identical to that of the Galactic globular cluster M92
with a relative age-difference uncertainty ranging from 10\% to 21\%.
The values for average aperture corrections,
$\langle \Delta_r \rangle$, are listed in
Table \ref{tbl-delta}\firstuse{Tab\ref{tbl-delta}}.
The zero-order (``breathing'') aperture corrections for these observations
($\delta_r$ : see
Table \ref{tbl-sdelta}\firstuse{Tab\ref{tbl-sdelta}})
were computed using a large aperture with a radius of 3.0 pixels
and a background annulus of
$3.0\leq r_{\rm sky}\leq6.5$ pixels.

Most of these observations were obtained when the WFPC2 CCDs operated at a
temperture of $-76\,^\circ$C
(the standard CCD operating temperature before 1994 April 23).
At this temperature, the number of ``hot''
pixels on a WFPC2 CCD would grow at a rate of several thousand pixels
per month per chip (Holtzman \et \cite{hoet1995a}).
There was a
small but statistically significant position shift for stars between the
F450W and F555W images.  Hot pixels and other CCD defects did not exhibit
this position shift.  We took advantage of this fact to reject all point-source
candidates with a position shift that were
not within 2 average deviations
of the median shift on each WFPC2 CCD.  This procedure allowed us to
statistically remove most of the hot pixels and other CCD defects.

We have used the standard Charge Transfer Effect correction
(4\% uniform wedge along the Y-axis of each CCD) recommended by
Holtzman \et (\cite{hoet1995b})
for all our observations regardless of the CCD operating temperature
at which they were obtained.
While Holtzman \et (\cite{hoet1995b}) consider $-76\,^\circ$C
data to have larger uncertainties associated with CTE, we note that
the application of the standard 4\% CTE correction worked well
in the Hodge 11 analysis of Mighell \et (\cite{miet1996}) whose
$V$ magnitude zeropoint differed from Walker's (\cite{wa1993})
by a statisically insignificant $0.009\,\pm\,0.010$ mag.
The Hodge 11 data analyzed by Mighell \et (\cite{miet1996})
was obtained on 1994 February 1;
the use of a CTE correction of 12\%
used by Holtzman \et (\cite{hoet1995b}) on their $-76\,^\circ$C data
would have clearly resulted in worse photometry for the Hodge 11 observations.

We present our WFPC2 stellar photometry of stars found in
NGC 416, NGC 121, NGC 339, NGC 361, and Kron 3 in
Tables
\ref{tbl-ngc416}\firstuse{Tab\ref{tbl-ngc416}},
\ref{tbl-ngc121}\firstuse{Tab\ref{tbl-ngc121}},
\ref{tbl-ngc339}\firstuse{Tab\ref{tbl-ngc339}},
\ref{tbl-ngc361}\firstuse{Tab\ref{tbl-ngc361}},
and
\ref{tbl-kron3}\firstuse{Tab\ref{tbl-kron3}},
respectively.
The first column gives the identification (ID) of the star.
The left-most digit of the ID gives the WFPC2 chip number (1, 2, 3, or 4)
where the star was found.
The right-most 4 digits gives the $x$ coordinate of the star multiplied
by 10.  The remaining 4 digits gives the $y$ coordinate of the star multiplied
by 10.
For example, the first star in Table \ref{tbl-ngc416} has an ID of
106084534
which indicates that it has the $(x,y)$ position of
$(60.8,453.4)$
on the PC1 CCD.  The second and third columns
give the $V$ magnitude and its rms ($1\,\sigma$) photometric error $\sigma_V$.
Likewise, the fourth and fifth columns give the $\bmv$ color and its
rms ($1\,\sigma$) photometric error $\sigma_{(B\!-V)}$.
We only present photometry of stars with signal-to-noise ratios
SNR$\,\geq\,$10 in both the F450W and F555W filters.

\section{COLOR-MAGNITUDE DIAGRAMS}

The $V$ versus $\bmv$ color-magnitude diagrams of our observed stellar
fields in
NGC 416, NGC 121, NGC 339, NGC 361, and Kron 3 are displayed in Figs.\
\ref{fig-ngc416}\firstuse{Fig\ref{fig-ngc416}},
\ref{fig-ngc121}\firstuse{Fig\ref{fig-ngc121}},
\ref{fig-ngc339}\firstuse{Fig\ref{fig-ngc339}},
\ref{fig-ngc361}\firstuse{Fig\ref{fig-ngc361}},
and
\ref{fig-kron3}\firstuse{Fig\ref{fig-kron3}},
respectively.
We have arbitrarily split each cluster observation into two regions:
(1) the ``cluster'' region
defined as the PC1 CCD,
and
(2) the SMC ``field'' region
defined as the WF CCDs.
Ideally, we should have used a background field that was
well outside the tidal radius of the cluster, however we have only
one WFPC2 field for each cluster.

We used the following procedure to statistically remove the SMC
field population from the cluster region CMDs.
For a given star in the cluster region CMD of NGC 416
(Fig.\ \ref{fig-ngc416}b)
we can count how many stars can be found in that CMD
that have $\bmv$ colors within
MAX$(2\sigma_{(B-V)},0.100)$ mag
and $V$ magnitudes within
MAX$(2\sigma_V,0.200)$ mag.
Let us call that number $N_{\rm NGC416}$.
We can also count how many
stars can be found in the {\em field} region CMD of NGC 416
(Fig.\ \ref{fig-ngc416}c)
within the same $V$ magnitude range and $\bmv$ color range
that was determined for the star in the {\em cluster} region CMD.
Let us call that number $N_{\rm SMC}$.
The probability, $p$,
that the star in the cluster region CMD
is actually a cluster member of
NGC 416 can be approximated as
\begin{equation}
p \approx 1 - {\rm MIN}
\left(
\frac{\alpha N_{\rm SMC}^{\rm UL84} }{N_{\rm NGC416}^{\rm LL95} }
,1.0
\right)~,
\label{eq:p}
\end{equation}
where
\begin{equation}
N_{\rm SMC}^{\rm UL84}
\approx
(N_{\rm SMC} + 1)
\left[
1
- \frac{1}{9(N_{\rm SMC} + 1)}
+ \frac{1.000}{3\sqrt{N_{\rm SMC} + 1}}
\right]^3
\label{eq:ul84}
\end{equation}
(Eq.\ 9 of Gehrels \cite{ge1986})
is the estimated upper $\sim$84\% confidence limit\footnote{
Corresponding to Gaussian statistics of $+1$ standard deviation:
$
\frac{1}{\sigma \sqrt{2\pi}}
\int_{-\infty}^{+1\sigma}
e^{\frac{-x^2}{2\sigma^2}}
\,dx
= 0.8413~.
$
}
of $N_{\rm SMC}$,
\begin{equation}
N_{\rm NGC416}^{\rm LL95}
\approx
N_{\rm NGC416}
\left[
1
- \frac{1}{9N_{\rm NGC416}}
- \frac{1.645}{3\sqrt{N_{\rm NGC416}}}
+ 0.031 N_{\rm NGC416}^{-2.50}
\right]^3
\label{eq:ll95}
\end{equation}
is the estimated lower 95\% confidence limit of $N_{\rm NGC416}$
(Eq.\ 14 of Gehrels \cite{ge1986}),
and
$\alpha\equiv0.0705$ which is the ratio of the area of
the cluster region
[$\sim$0.314 arcmin$^2$]
to the area of the LMC field region
[$\sim$4.45 arcmin$^2$].
The probable cluster membership of any given star can be estimated
by picking a uniform random number, $0 \leq p^\prime \leq 1$,
and if $p^\prime \leq p$ then
the star is said to be a probable cluster member.

We demonstrate this CMD-cleaning method using the first star
in Table \ref{tbl-ngc416} which has a $V$ magnitude of
$19.701$$\pm$$0.011$ and a
$\bmv$ color of $0.796$$\pm$$0.017$ mag.
We found
$N_{\rm NGC416} = 170$ stars on the PC1 CCD (see Fig.\ \ref{fig-ngc416}b)
with $V$ magnitudes between
$19.501$ and $19.901$
and $\bmv$ colors between
$0.696$ and $0.896$ mag.
Similarly, we found
$N_{\rm SMC} = 130$ stars on the WF CCDs (see Fig.\ \ref{fig-ngc416}c)
within the same $V$ magnitude range and $\bmv$ color range.
{}From Eq.\ \ref{eq:ul84} we find that
$N_{\rm SMC}^{\rm UL84} \approx 142.43$ stars
and from Eq.\ \ref{eq:ll95} we find that
$N_{\rm NGC416}^{\rm LL95} \approx 149.14$ stars.
Equation \ref{eq:p} thus gives the
probability that this star belongs to the cluster NGC 416
as $p \approx 0.9327$.
Our uniform random number generator gave us the value $p^\prime = 0.1384$.
Since $p^\prime \leq p$,
we claim that this star is a probable cluster member of NGC 416.
Based on its position in the color-magnitude diagram of
Fig.\ \ref{fig-ngc416}d, we see that
this star is probably a helium core-burning horizontal-branch star.

Following the methodology described above, we
determined the probable cluster membership for all 3351 stars in
the cluster region CMD field of NGC 416
using a uniform random number generator.
A total of 2826 stars were found to be
probable cluster members and they are displayed in the cleaned
cluster CMD (see Fig.\ \ref{fig-ngc416}d).
This CMD-cleaning method is probabilistic and
{Fig.\ \ref{fig-ngc416}d} therefore
represents only one out of an infinite number of different possible
realizations of the cleaned NGC 416 CMD.

Equation \ref{eq:p} was designed to eliminate the strong contamination
of young SMC field stars that are clearly seen in the NGC 416 observation.
Unfortunately, the clear presence of cluster stars on the WF CCDs (see Fig.\ 1)
indicates that this cleaning procedure represents an ``overcleaning''
of the true cluster CMDs.  Such overcleaning will
be most obvious in regions of the color-magnitude diagram where
there is rapid luminosity/color evolution or where stellar densities
are intrinsically low.
An example of both effects can be found in
the cleaned cluster CMD of NGC 339 (Fig.\ \ref{fig-ngc339}d) where
a large gap is clearly seen in the red giant branch just below the
horizontal branch.
In as much as most of the contaminating SMC field
main-sequence stars have been removed
from the resulting cleaned CMDs, we see that Eq.\ \ref{eq:p} has achieved its
design goal.  We caution the reader against placing excessive reliance
on the actual cluster membership of any particular star in the cleaned
CMDs.

\section{REDDENINGS AND METALLICITIES}

\subsection{Simultaneous Reddening and Metallicity Method}

The work of Sarajedini (\cite{sa1994}) introduced a method by which
the the reddening and metallicity of a globular cluster can be determined
simultaneously using a calibration based on several well-observed Galactic
globular clusters with abundances measured on the Zinn \& West
(\cite{ziwe1984})
metallicity scale. All that is required are values for the
magnitude level of the horizontal branch (HB),
the color of the red giant branch (RGB) at the level of the HB,
and the shape and the position of the RGB.
The simultaneous reddening and metallicity (SRM)
method has subsequently been developed for use with several standard
filter combinations\footnote{Various forms of the SRM method:
$V$ \vs $V\!-\!I$ (Sarajedini \cite{sa1994}),
$R$ \vs $B\!-\!R$ (Sarajedini \& Geisler \cite{sage1996}),
$V$ \vs $B\!-\!V$  (Sarajedini \& Layden \cite{sala1997}),
$T_1$ \vs $C-T_1$ (Geisler \& Sarajedini \cite{gesa1998}).},
and we would like to apply the $V$ \vs $B\!-\!V$ version to the photometry
presented herein. One might wonder if there
is some danger in applying the SRM method to clusters with ages that are much
younger than those of the Galactic globular clusters.
Sarajedini \& Layden (\cite{sala1997}) discuss the age
sensitivity of the SRM method. They conclude that the method is largely
insensitive to age effects for red HB clusters older than $\sim$5 Gyr. As we
shall see, this applies to all of the clusters in the present work. We
acknowledge
however that there {\it might} be small systematic errors remaining in our
abundance measurements.

The first step in applying the SRM method is to measure the $V$ magnitude of
the
red HB clump ($V_{\rm{RHB}}$).
To do this, we follow the procedure described by
Sarajedini, Lee, \& Lee (\cite{salele1995}; hereafter SLL).
Figure \ref{rgb_fits}
\firstuse{Fig\ref{rgb_fits}}
shows the CMDs of the cluster RGBs and HBs,
where we have included existing CCD photometry for Lindsay 113
(Mould \et \cite{moet1984}, see also {\sc{PaperI}}\,\footnote{
The Lindsay 113 $BR$ photometry of Mould \et (\protect\cite{moet1984})
was converted
by us in {\sc{PaperI}} to the $BV$ system using the color transformation
equation of Sarajedini \& Geisler (\cite{sage1996}):
$
(B\!-\!V)_0
=
0.01135
+ 0.6184(\!B\!-\!R)_0
- 0.1071(\!B\!-\!R)_0^2
+ 0.1249(\!B\!-\!R)_0^3
- 0.0319(\!B\!-\!R)_0^4
$.
This transformation equation has an rms error of 0.018 mag and was
derived by analyzing bright stars from many
globular clusters over a wide range of metallicities spanning a color range
$-0.2 \leq (\!B\!-\!R) \leq 2.5$ mag. The addition of the new photometry from
von Braun et al. (\cite{voet1998}), which includes the stars in
4 globular clusters along with the super-metal-rich open cluster NGC 6791,
only slightly increases the rms error to 0.019 mag.
} 
\,).  The rectangles indicate the stars
used in the determination of
$V_{\rm{RHB}}$, which are listed in
Table \ref{tbl-clpar}\firstuse{Tab\ref{tbl-clpar}}.
The error in this quantity is
computed by combining, in quadrature, the standard error of the mean
with the estimated error in the photometric zeropoint, which
is $\pm$0.05 mag for the {\em{HST}} observations and $\pm$0.02 mag for the
ground-based observations. The agreement between the $V_{\rm{RHB}}$ values for
the HST cluster observations and existing ground-based measurements for
these same clusters is quite reasonable. For example, converting
the Kron 3 $B-R$ data of Rich \et (\cite{riet1984}) to $B-V$ yields a
$V_{\rm{RHB}}$ value of 19.44
(DH98)
in excellent agreement with $V_{\rm{RHB}}$=19.45 from our HST data.
We can perform the
same experiment with the NGC 121 $B-R$ data of Stryker \et (\cite{stet1985});
this gives $V_{\rm{RHB}}$=19.66 (albeit for only 7 stars) which agrees with
our value to within the errors. Lastly, DH98
quote $V_{\rm{RHB}}$=19.36 for NGC 339 based on some unpublished photometry;
this is close to our value of 19.46 for this cluster.

The next step is the determination of RGB
fiducials for each cluster. This is accomplished via an iterative
2$\sigma$ rejection polynomial fitting procedure applied to the RGB
stars. The resultant fits are plotted in
Fig. \ref{rgb_fits}.
The analogous diagram for Lindsay 1 is given in Fig.\ 1 of SLL.
{}From these fits, we
measure the value of
$(B\!-\!V)_g$
which is the $B\!-\!V$ color of the RGB at the level of the HB.
All of the inputs into the SRM method are now in hand and
it is a simple matter to apply the $V$ \vs $B\!-\!V$ version
to our 7 SMC clusters.
We also have estimated errors for $V_{\rm{RHB}}$ and $(B\!-\!V)_g$. These are
input into the SRM method via a Monte Carlo simulation which yields
errors in $\feh$ and $\ebmv$ (see Sarajedini \cite{sa1994} for
details).

\subsection{Red Giant Branch Slope}

Another method we can exploit for the determination of metallicity
is the RGB slope.
It is well known that
the RGB slope steepens with decreasing metallicity.
To establish a modern calibration of how metallicity
varies with RGB slope, we turn to the photometric data presented by
Sarajedini \& Layden (\cite{sala1997}).
In particular, we measure the
slopes of the RGBs for all clusters listed in their Table 5.
We follow Hartwick (\cite{ha1968}) and define the
RGB slopes ($S_{-2.0}$ and $S_{-2.5}$) as follows:
$
S_{-2.0}
\equiv
-2.0
/
\left[\,\left(B\!-\!V\right)_g - \left(B\!-\!V\right)_{-2.0}\,\right]
$
and
$
S_{-2.5}
\equiv
-2.5
/
\left[\,\left(B\!-\!V\right)_g - \left(B\!-\!V\right)_{-2.5}\,\right]
$
where we measure the color of the RGB at 2.0 and 2.5 magnitudes above the
HB. Table \ref{tbl-slopes}\firstuse{Tab\ref{tbl-slopes}}
lists the calibration data measured from the RGBs of
Sarajedini \& Layden (\cite{sala1997}) along with the cluster metallicities
which are on the Zinn \& West (\cite{ziwe1984}) scale.
The clusters listed first are
their ``primary'' calibrators and those listed second are their
``secondary'' calibrators. The two panels of
Fig. \ref{fig-slopes}\firstuse{Fig\ref{fig-slopes}}
show our weighted
least squares fits to the data. The $S_{-2.0}$ fit has an rms error of 0.19 dex
while the $S_{-2.5}$ fit exhibits an rms error of 0.14 dex.
We note that, although Lindsay 1 is listed in Table \ref{rgb_fits},
it was not used in the fit because
we seek to determine its metallicity via the derived relation.

Once we have established the relations between metallicity and RGB slope,
it is a simple matter to measure the slopes for our 7 SMC clusters
from the RGB fits. For the
two clusters with RGBs that reliably extend 2.5 magnitudes above the
HB (NGC 121 and NGC 416), we use the $S_{-2.5}$ calibration; for the others,
we apply the $S_{-2.0}$ relation. In all cases, the derived metallicities
from the RGB slope are within 0.05 dex of the SRM method determination,
lending credence to our techniques.
The RGB slope metallicity can then be combined with Equation 1 of
Sarajedini \& Layden (\cite{sala1997}) and the measured value of
$(B\!-\!V)_g$ to compute the reddening. Again, these reddenings are
within 0.01 mag of those yielded by the SRM method.

To arrive at our final adopted values, we perform a weighted mean of
the SRM method and the RGB slope results. This process gives us the
metallicities and reddenings tabulated in
Table \ref{tbl-clpar}. A careful inspection of
Table \ref{tbl-clpar} shows that the reddening of Kron 3 is a negative value.
This is, of course, not possible, but given the error in the
reddening of $\pm$0.02 mag, it is, within the errors, consistent with
$\ebmv=0$.

Recently, Da Costa \& Hatzidimitriou (DH98)
have published
\ion{Ca}{2} triplet metallicities for 5 of the clusters in our sample
(Lindsay 1, Kron 3, NGC 121, NGC 339, and Lindsay 113). If we compare our
abundances for these clusters with their values, we find a mean difference
of $\feh_{\rm{CaII}} - \feh = 0.11 \pm 0.06$ (s.e.m.) when comparing to their
metallicities which have not been corrected for age effects (column 2 of Table
3 in
Da Costa \& Hatzidimitriou)
and $0.06 \pm 0.08$ (s.e.m.) for their age-corrected metallicities (column 3 of
Table 3). Note that DH98 utilized the Lindsay 1
photometry of Olszewski \et (\cite{olet1987}) in their spectroscopic analysis.
In their paper, Olszewski \et (\cite{olet1987}) construct a CMD of Lindsay 1
and estimate the V magnitude of
the red HB clump to be $19.2\pm0.1$, which DH98 adopt in their work.
However, an examination of the Lindsay 1 CMD reveals that this is
{\it clearly} too bright by $\sim$0.15 mag. Our value for the $V_{\rm{RHB}}$ of
Lindsay 1, which has been determined from the
Olszewski \et (\cite{olet1987}) data (see SLL), is more faithful to the actual
location
of this feature. If we correct the DH98 Zinn \& West metallicity of Lindsay 1
using the
$V_{\rm{RHB}}$ value in Table \ref{tbl-clpar}, we find a decrease of 0.05 dex
bringing it closer to our metallicity value. Furthermore, if we adopt a more
appropriate age for Kron 3 (see next section), the DH98 age-corrected abundance
becomes more metal-poor by $\sim$0.03 dex, again closer to our value.
Taken together, the good agreement between the spectroscopic
metallicities and our photometric ones provides a useful check on
both methods. Furthermore, it indicates that any systematic errors still
present in our metallicities are likely to be small.

\section{AGES}

The most robust age determination techniques are those that deal with the
measurement of relative ages. As such, we have chosen to study the
ages of the SMC populous clusters relative to that of Lindsay 1.
One would ideally like to utilize the position
of the main sequence turnoff (MSTO) to estimate the relative age. However, in
the present case, the scatter of the main sequence photometry prohibits
us from measuring the precise location of the MSTO. As a result, we
must resort to a less direct technique to determine the cluster age.
In particular, since our photometry reveals the location of the red
HB clump very well, we can utilize the age-determination method described
by SLL. They exploit the fact that the color of the red HB clump is
dependent on metallicity and age in order to derive the ages of several
Galactic globular clusters. After measuring the difference in $B\!-\!V$ color
between the HB and RGB [$\dbmv$] and adopting values for the cluster
metallicity, SLL use their Fig.\ 4 to compute the cluster ages. Their
Fig. 5 then yields the absolute magnitude of the red HB given the
age and metallicity. Both of these calibrations are based on theoretical
synthetic HB models. Sarajedini \et (\cite{salele1995}) then test
their $\dbmv$ relative ages by comparing them to those yielded by
MSTO comparisons. They find that for clusters with $\feh < -0.7$ dex,
the $\dbmv$ ages and the MSTO ages agree to within 1 Gyr. We also
point out that, since $\dbmv$ is a purely differential quantity,
it is relatively easy to measure precisely, and it is free from the
uncertainties inherent in the photometric zeropoint; it is also
reddening and distance independent.

To apply the $\dbmv$ method of SLL, we must measure the color of the
red HB clump and the RGB at the level of the HB. The latter has already been
described in previous section.
The color of the red HB is simply computed using
the stars in the rectangles illustrated in Fig. \ref{rgb_fits}.
The associated errors
represent the standard error of the mean. This procedure gives us the
$\dbmv$ values listed in Table \ref{tbl-clpar}.
Figure 4 of SLL then
yields the ages tabulated in column 6 of Table \ref{tbl-clpar}.
The absolute $V$ magnitude of the RHB,
$M_V^{\rm{RHB}}$,
is then derived from Fig.\ 5 of SLL and listed
in column 7 of Table \ref{tbl-clpar}.
We would like to tie our ages to the scale of
Olszewski \et (\cite{olet1996}) in which Lindsay 1 is $9\pm1$ Gyr old.
As a result, we have added $1.3\pm1.1$ Gyr to the $\dbmv$ ages derived above.
Our adopted ages are those surrounded by brackets in
Table \ref{tbl-clpar}.
These ages supersede those published in {\sc{PaperI}}.

We note that, since the youngest age in the calibration of SLL is 7 Gyr,
we have had to extrapolate their model grid in order to estimate the ages of
Lindsay 113, Kron 3, NGC 339, and NGC 416. We realize that this is not
ideal; however, we have no choice given the large errors associated with the
main sequence photometry.
For the sake of completeness, we note that the extrapolations have been
performed using cubic polynomials fitted to the 9 age points (7 to 15 Gyr in
units
of 1 Gyr) in Fig. 4 of SLL. The root-mean-square (rms) deviations of the fitted
points from the fits were never greater than 0.1 Gyr with typical values being
$\sim$0.06 Gyr. Because these rms deviations are so small, they were not
included
in the derived age error.

One way we can verify the resulting ages is to compare the
CMDs with theoretical isochrones. This is not the preferred manner in which to
measure
these ages because of lingering uncertainties in the models, but it
does provide a quantitative check on our results.
Figures \ref{fig-iso}\firstuse{Fig\ref{fig-iso}}
and
\ref{fig-iso2}\firstuse{Fig\ref{fig-iso2}}
show the comparisons of the cluster photometry with
the theoretical isochrones of Bertelli \et (\cite{beet1994}) for the
indicated ages and metallicities. Figure \ref{fig-iso} displays the three
clusters whose metallicities are close to those for
which Bertelli \et (\cite{beet1994}) tabulate isochrones thus requiring
essentially
no interpolation within their (coarse) metallicity grid. Figure \ref{fig-iso2}
includes the remaining clusters which require comparisons to isochrones with
metallicities that bracket the measured cluster value. The isochrones have been
adjusted in the horizontal direction using the reddenings in Table
\ref{tbl-clpar}
and in the vertical direction by requiring a match between the observed
$V_{\rm{RHB}}$ values and the theoretical location of the red HB.
It is evident from
Figs. \ref{fig-iso} and \ref{fig-iso2} that the rescaled $\dbmv$ ages of these
clusters (see bracketed values of column 6 of Table \ref{tbl-clpar})
are fully consistent with the isochrone fits.
This consistency is one piece of evidence that our
relative
$\dbmv$ age estimates are robust, even though they are extrapolations.

Another method at our disposal to verify the $\dbmv$ ages involves
direct comparisons between the photometry for each cluster and the
fiducial sequence of a comparison cluster. We seek to examine the relative
locations of the MSTOs in the same spirit as SLL.
Figure \ref{fig-cmds}\firstuse{Fig\ref{fig-cmds}}
illustrates our MSTO comparisons.
The strategy is to use the $M_V^{\rm{RHB}}$ and $\ebmv$ values in
Table \ref{tbl-clpar} to place the cluster photometry
into the Hertzprung-Russell diagram;
then we would like to compare
each cluster to the fiducial sequence of a
standard cluster of similar metallicity.
The choice of the standard cluster is obvious given the fact that we
have selected Lindsay 1 to set our age scale. However, the metallicity of
NGC 121 is too different from that of Lindsay 1 for the latter to serve
as an effective standard. As a result, we will utilize the fiducial sequence
of Palomar 14 ($\feh = -1.60$) from Sarajedini (\cite{sa1997}) to compare
to NGC 121. The clusters are presented in Fig. \ref{fig-cmds}
in order of increasing age from top to bottom and left to right.
The primary result of Fig. \ref{fig-cmds}
is that the age ranking derived from the $\dbmv$ method is corroborated
by the relative locations of the MSTOs. Notice, for example, the locations
of the subgiant branches (i.e. the nearly horizontal sequence that links
the MSTO and RGB); the magnitude of the SGBs approaches that of the
standard cluster fiducials as age increases. Another feature to notice about
Fig. \ref{fig-cmds}
is that the RGBs of all the clusters line up quite well with that
of the standard cluster, thus supporting our metallicity measurements.
In addition, we note that the color of the red HB clump becomes
progressively bluer for older clusters, as it should, an effect also
pointed out in the review article by Sarajedini et al. (\cite{sachde1997}).

There are existing age determinations for some of the SMC clusters
presented herein. For example, Mould \et (\cite{moet1984}) utilize their
$B-R$ CCD photometry to estimate an age of 5 Gyr for Lindsay 113 when placed
at a distance modulus of $(m-M)_0 = 18.80$ and $[M/H] = -1.4$ is adopted.
Using similar parameters, Olszewski \et (\cite{olet1987}) find an age of 10 Gyr
for Lindsay 1, while Stryker \et (\cite{stet1985}) conclude that NGC 121 is
12 Gyr old. All three of these ages are in excellent agreement with those
estimated in this study. In the case of Lindsay 113, this is not
unexpected since the photometry we use to derive an age is simply that of
Mould \et (\cite{moet1984}) converted from $B-R$ to $B-V$ (see Sec. 4.1).
For Kron 3, the work of Rich \et (\cite{riet1984}) yields an age between
5 and 8 Gyr; in contrast, Alcaino \et (\cite{alet1996}) conclude that
the age of Kron 3 is 10 Gyr with a lower limit of 8 Gyr.
The recent work of DH98 concerning the \ion{Ca}{2} triplet metallicities
of these clusters has adopted 9 Gyr for the age of Kron 3. Based on the
relative
location of the red HB clump as well as the MSTO, we have shown in
this paper that the age of Kron 3 is closer to 6 Gyr, thus corroborating the
original age estimate of Rich \et (\cite{riet1984}). When we compare the
Kron 3 CMD of Alcaino \et (\cite{alet1996}) with our photometry, we find
generally good agreement in terms of the magnitude of the MSTO. As a result,
it is unclear to us at this point why Alcaino \et (\cite{alet1996}) derive
a much older age for Kron 3.

\newpage
\section{DISCUSSION}

The exhaustive reduction and analysis presented so far will now provide
the foundation for the remainder of the paper in which we hope to shed
light on the star formation history (SFH) of
the SMC. To facilitate this,
Fig. \ref{fig-feh_vs_age}\firstuse{Fig\ref{fig-feh_vs_age}}
shows the relationship between
age and metallicity for the SMC clusters
(filled circles) considered herein.
The open circles are the
younger SMC clusters taken from the discussion presented by DH98 (see DH98 for
references). The remaining three points, the asterisk, diamond, and square,
are the present-day abundance of the SMC taken from
Luck \& Lambert (\cite{lula1992}), Russell \& Bessell (\cite{rube1989}), and
Hill (\cite{hi1997}), respectively.
The points marked as {\protect\large$\times$} are LMC clusters;
the data for clusters older than 10 Gyr are from
Olsen \et (\cite{olet1998}) and the younger cluster data are from
Geisler \et (\cite{geet1997})
and Bica \et (\cite{biet1998}).
{}From the appearance of this
figure, we note the following points. First, assuming that cluster destruction
processes and cluster fading with age are similar in the LMC and SMC,
it is clear that, in general,
these two galaxies have had very different SFHs.
In particular, at a given age,
the mean abundance
of the SMC clusters is $\sim$0.3 dex lower than that of the LMC clusters
(Olszewski et al. \cite{olet1996}); this is a restatement of the observation
that for dwarf galaxies, the mean metal abundance is correlated with
the total absolute luminosity in the sense that more metal-rich systems are
brighter (Sarajedini \et \cite{saet1997}, and references therein).
However, Fig. \ref{fig-feh_vs_age} allows us to draw more specific conclusions
than this. For example, when considering
only the clusters with ages less than $\sim$10 Gyr, it seems that between
$\sim$3
and $\sim$10 Gyr ago, the LMC experienced a rapid
chemical enrichment (Geha \et \cite{geet1998}) that the SMC did not,
or perhaps the chemical enrichment of the LMC was simply more vigorous
than that of the SMC (Pagel \& Tautvai\u{s}ien\.{e} \cite{pata1998}).

The star formation history derived from the LMC clusters differs
markedly from that indicated by the field stars (Geha \et \cite{geet1998};
Olszewski \et \cite{olet1996}; Sarajedini \cite{sa1998}).
There is tenuous evidence that,
in the case of the SMC, the SFHs of the clusters and field stars
maybe quite similar
(e.g.\ DH98;
Gardiner \& Hatzidimitriou \cite{gaha1992}).
Sarajedini (\cite{sa1998}) argues that this apparent difference between
the LMC and the SMC
is due to the fact that there are probably more clusters in the LMC
age gap (between 2.5 and 9 Gyr) than are currently known.
Much work has been done on the young field stars in the Magellanic
Clouds, however
progress towards understanding the early SFH of the Clouds
can only be achieved by analyzing the old (i.e.\ faint)
field stellar population.
While the analysis of a few {\em{HST}} WFPC2 fields
(e.g. Gallagher \et \cite{gaet1996};
Holtzman \et \cite{hoet1997};
Geha \et \cite{geet1998})
has shown the clear potential of such research for understanding
of the early SFH of the LMC,
comparable studies in the SMC do not yet exist.

Figure \ref{fig-amr_models}\firstuse{Fig\ref{fig-amr_models}}
compares our age-metallicity data for the SMC with two
theoretical representations of its SFH. Under the assumption of
chemical homogeneity, the top panel shows the result
of treating the SMC as a simple closed-box system with continuous
star formation, kindly provided by Gary Da Costa  and illustrated
in Fig.\ 4 of DH98, whereas
the lower panel depicts the bursting SFH of Pagel
\& Tautvai\u{s}ien\.{e} (\cite{pata1998}). The SMC is almost certainly
not a closed-box, but comparisons such as these
allow us to distinguish between a continuous and bursting SFH.
Indeed, the upper
panel showing the continuous star formation is an exceedingly poor
fit to the age-abundance data. Of the older clusters considered here,
only the error bars associated with NGC 121 and perhaps Lindsay 1 are
consistent with this curve. In contrast, the theoretical relation
shown in the bottom panel is a better fit to the observational data.
The majority of the clusters intersect
the solid curve to within their 1$\sigma$ error bars. The three possible
exceptions are NGC 416, NGC 361, and NGC 339.
The bursting star formation model formulated by Pagel
\& Tautvai\u{s}ien\.{e} (\cite{pata1998}) can be adjusted to fit
the age-metallicity data for the SMC. For example, the existing
form of their model postulates an initial star formation burst that
began 14 Gyr ago and peaked $\sim$11.3 Gyr ago. Then the star formation
rate (SFR) decreased and remained constant until 4 Gyr ago
(see their Fig.\ 2).
If we adjust their model so that
the initial burst of star formation lasted for only 2 Gyr as opposed to
2.7 Gyr, then it is possible that the age-metallicity relation
would provide a better fit to all of the older clusters. With this one
revision, the SFH outlined by Pagel \& Tautvai\u{s}ien\.{e}
(\cite{pata1998}) would be an excellent fit to the age-abundance data for
the clusters in the SMC.

\section{CONCLUSIONS AND FUTURE WORK}

We have presented new {\em{HST}} WFPC2 photometry
for 5 star clusters in the Small Magellanic Cloud.
Along with the published data for two other clusters,
we derive more precise values for the cluster metallicities and ages
as compared with previous studies. Knowledge of these quantities has
allowed us to improve our understanding of the age-metallicity relation
of the SMC clusters, and provided, for the first time, a consistent picture
for the star formation and chemical enrichment history of the SMC.

Figure \ref{fig-clusters_field}\firstuse{Fig\ref{fig-clusters_field}}
compares the probable cluster members with the nearby SMC field stellar
population.
While the cluster/field separation scheme used in this paper could
be improved, one can nevertheless draw several conclusions from this
figure which might be helpful to researchers designing future
studies of the populous star clusters in the Small Magellanic Cloud.
First, the youngest stars in the SMC field stellar population near the
populous clusters cover a large age range of over 10 Gyr.
The bright blue main-sequence stars of
Fig.\ \ref{fig-ngc416}c
indicate that the SMC field stellar population
near NGC 416 is young ( $<$$1$ Gyr)
while the age of the SMC field near Kron 3 and NGC 121 is
$7$$\pm$$2.5$
and
$11$$\pm$$2.5$ Gyr,
respectively
(Gardiner \& Hatzidimitriou \cite{gaha1992}).
Second, while the red giant branches of the field and cluster populations
are nearly identical, one sees many
field subgiant branch (SGB) stars which are significantly brighter
than the average cluster subgiant branch star at the same $B\!-\!V$ color.
This is consistent with the interpretation that the SMC field stellar
population near the populous clusters have
intermediate-age metal-poor components which are slightly
younger than the nearby cluster but of similar metallicity
(assuming the distance to the cluster and field are the same).
Third, the spread of $B\!-\!V$ color
near the base of the red giant branch is anomalously large
for some of the cluster CMDs (e.g.\ NGC 121).
This is most likely a result of
inadequate CMD cleaning due to poor sampling of the SMC field subgiant branch
stellar population.
Better sampling of the SMC field SGB stars is required
for the identification of probable field SGB stars.
Since SGB stars are expected to be
relatively rare in any given observation of the SMC field,
due to their rapid luminosity and color evolution,
one must then observe a lot of the nearby SMC field in order to adequately
determine the luminosity and color properties of the field SGB population.
In other words, the accurate determination of the age of an SMC populous
cluster via the analysis of main-sequence-turnoff photometry
requires the proper identification of probable SMC field SGB stars
--- one WFPC2 field-of-view is clearly inadequate for some clusters.
Deeper observations with better field coverage are thus
required in order to obtain unambiguous astrophysical interpretations of the
cluster/field star formation histories of the Small Magellanic Cloud.

\acknowledgments

We thank Gary Da Costa for the timely provision of his SMC star formation
history models.
We would like to thank the anonymous referee whose comments and suggestions
have improved the final product presented here.
KJM
was supported by a grant from
the National Aeronautics and Space Administration (NASA),
Order No.\ S-67046-F, which was awarded by
the Long-Term Space Astrophysics Program (NRA 95-OSS-16).
AS
was supported by the
NASA
grant number HF-01077.01-94A from
the Space Telescope Science
Institute, which is operated by the Association of Universities for
Research in Astronomy, Inc., under NASA contract NAS5-26555.
AS wishes to thank Lick Observatory for their generosity and
hospitality during his visit.
Figure \protect\ref{fig-1}\ was created with images from the
Digitized Sky Survey\footnote{
Based on photographic data obtained using The UK Schmidt Telescope.
The UK Schmidt Telescope was operated by the Royal Observatory
Edinburgh, with funding from the UK Science and Engineering Research
Council, until 1988 June, and thereafter by the Anglo-Australian
Observatory.  Original plate material is copyright (c) the Royal
Observatory Edinburgh and the Anglo-Australian Observatory.  The
plates were processed into the present compressed digital form with
their permission.  The Digitized Sky Survey was produced at the Space
Telescope Science Institute under US Government grant NAG W-2166.
}.
This research has made use of
NASA's Astrophysics Data System Abstract Service
and the NASA/IPAC Extragalactic Database (NED)
which is operated by the Jet Propulsion Laboratory at the
California Institute of Technology, under
contract with NASA.

\newpage
\begin{table}
\dummytable\label{tbl-obslog}
\ifundefined{showtables}{
}\else{
  \vspace*{-2cm}
  \hspace*{-20mm}
  \epsfxsize=8.0truein
  \epsfbox{mighell.tab01.eps}
}
\fi
\end{table}

\newpage
\begin{table}
\dummytable\label{tbl-delta}
\ifundefined{showtables}{
}\else{
  \vspace*{-2cm}
  \hspace*{-20mm}
  \epsfxsize=8.0truein
  \epsfbox{mighell.tab02.eps}
}
\fi
\end{table}

\newpage
\begin{table}
\dummytable\label{tbl-sdelta}
\ifundefined{showtables}{
}\else{
  \vspace*{-2cm}
  \hspace*{-20mm}
  \epsfxsize=8.0truein
  \epsfbox{mighell.tab03.eps}
}
\fi
\end{table}

\newpage
\begin{table}
\dummytable\label{tbl-ngc416}
\ifundefined{showtables}{
}\else{
  \vspace*{-2cm}
  \hspace*{-20mm}
  \epsfxsize=8.0truein
  \epsfbox{mighell.tab04.eps}
}
\fi
\end{table}

\newpage
\begin{table}
\dummytable\label{tbl-ngc121}
\ifundefined{showtables}{
}\else{
  \vspace*{-2cm}
  \hspace*{-20mm}
  \epsfxsize=8.0truein
  \epsfbox{mighell.tab05.eps}
}
\fi
\end{table}

\newpage
\begin{table}
\dummytable\label{tbl-ngc339}
\ifundefined{showtables}{
}\else{
  \vspace*{-2cm}
  \hspace*{-20mm}
  \epsfxsize=8.0truein
  \epsfbox{mighell.tab06.eps}
}
\fi
\end{table}

\newpage
\begin{table}
\dummytable\label{tbl-ngc361}
\ifundefined{showtables}{
}\else{
  \vspace*{-2cm}
  \hspace*{-20mm}
  \epsfxsize=8.0truein
  \epsfbox{mighell.tab07.eps}
}
\fi
\end{table}

\newpage
\begin{table}
\dummytable\label{tbl-kron3}
\ifundefined{showtables}{
}\else{
  \vspace*{-2cm}
  \hspace*{-20mm}
  \epsfxsize=8.0truein
  \epsfbox{mighell.tab08.eps}
}
\fi
\end{table}

\newpage
\begin{table}
\dummytable\label{tbl-clpar}
\ifundefined{showtables}{
}\else{
  \vspace*{-2cm}
  \hspace*{-20mm}
  \epsfxsize=8.0truein
  \epsfbox{mighell.tab09.eps}
}
\fi
\end{table}

\newpage
\begin{table}
\dummytable\label{tbl-slopes}
\ifundefined{showtables}{
}\else{
  \vspace*{-2cm}
  \hspace*{-20mm}
  \epsfxsize=8.0truein
  \epsfbox{mighell.tab10.eps}
}
\fi
\end{table}

\begin{table}
\dummytable\label{tbl-x}
\end{table}

\def\fig1cap{
\label{fig-1}
Digitized Sky Survey images showing the observed populous clusters
in the Small Magellanic Clouds:
NGC 121,
NGC 339,
NGC 361,
NGC 416,
and
Kron 3.
The outlines indicate the measured field-of-view of the
{\em{Hubble Space Telescope}} WFPC2 cameras at the four
target positions (see Table~\protect\ref{tbl-obslog}).
Each subfield shown subtends 5\arcmin\ on a side.
The orientation is North to the top and East to the left.
}

\ifundefined{showfigs}{
  \newpage
  \centerline{{\Large\bf{Figure Captions}}}
  \smallskip
  \figcaption[]{\fig1cap}
}\else{
  \clearpage
  \newpage
  \begin{figure}[p]
    \figurenum{1}
    \vskip -2mm
    \epsscale{0.40}
    \plotone{mighell.fig01.ngc121.eps}
    \plotone{mighell.fig01.ngc339.eps}\\
    \plotone{mighell.fig01.ngc361.eps}
    \plotone{mighell.fig01.ngc416.eps}\\
    \hskip -66truemm
    \plotone{mighell.fig01.kron3.eps}
    \epsscale{0.95}
    \vskip -2mm
    \caption[]{\baselineskip 1.15em \fig1cap}
  \end{figure}
}
\fi

\def\fig2cap{
\label{fig-ngc416}
The $V$ \vs $\bmv$ color-magnitude diagram of the observed stellar
field in the SMC populous cluster NGC 416.
{\bf{(a)}} The 8513 stars with
signal-to-noise ratios S/N$\geq$10 in
both filters are plotted (dots) along with the
2226 CCD/image defects (open circles).
{\bf{(b)}} The 3351 stars found on the PC1 CCD.
{\bf{(c)}} The 5162 stars found on the WF2, WF3, and WF4 CCDs.
{\bf{(d)}} The ``cleaned'' color-magnitude diagram of NGC 416
contains 2826 stars.
The error bars indicate rms ($1\sigma$)
uncertainties for a single star at the corresponding magnitude.
}

\ifundefined{showfigs}{
  \figcaption[]{\fig2cap}
}\else{
  \clearpage
  \newpage
  \begin{figure}[p]
    \figurenum{2}
    \vspace*{-2cm}
    \hspace*{-5mm}
    \epsfxsize=7.0truein
    \epsfbox{mighell.fig02.eps}
    \vspace*{-4cm}
    \caption[]{\baselineskip 1.15em \fig2cap}
  \end{figure}
}
\fi

\def\fig3cap{
\label{fig-ngc121}
The $V$ \vs $\bmv$ color-magnitude diagram of the observed stellar
field in the SMC populous cluster NGC 121.
{\bf{(a)}} The 8133 stars with
signal-to-noise ratios S/N$\geq$10 in
both filters are plotted (dots) along with the
2128 CCD/image defects (open circles).
{\bf{(b)}} The 4071 stars found on the PC1 CCD.
{\bf{(c)}} The 4062 stars found on the WF2, WF3, and WF4 CCDs.
{\bf{(d)}} The ``cleaned'' color-magnitude diagram of NGC 121
contains 3696 stars.
The error bars indicate rms ($1\sigma$)
uncertainties for a single star at the corresponding magnitude.
}

\ifundefined{showfigs}{
  \figcaption[]{\fig3cap}
}\else{
  \clearpage
  \newpage
  \begin{figure}[p]
    \figurenum{3}
    \vspace*{-2cm}
    \hspace*{-5mm}
    \epsfxsize=7.0truein
    \epsfbox{mighell.fig03.eps}
    \vspace*{-4cm}
    \caption[]{\baselineskip 1.15em \fig3cap}
  \end{figure}
}
\fi

\def\fig4cap{
\label{fig-ngc339}
The $V$ \vs $\bmv$ color-magnitude diagram of the observed stellar
field in the SMC populous cluster NGC 339.
{\bf{(a)}} The 5245 stars with
signal-to-noise ratios S/N$\geq$10 in
both filters are plotted (dots) along with the
1377 CCD/image defects (open circles).
{\bf{(b)}} The 1199 stars found on the PC1 CCD.
{\bf{(c)}} The 4046 stars found on the WF2, WF3, and WF4 CCDs.
{\bf{(d)}} The ``cleaned'' color-magnitude diagram of NGC 339
contains 773 stars.
The error bars indicate rms ($1\sigma$)
uncertainties for a single star at the corresponding magnitude.
}

\ifundefined{showfigs}{
  \figcaption[]{\fig4cap}
}\else{
  \clearpage
  \newpage
  \begin{figure}[p]
    \figurenum{4}
    \vspace*{-2cm}
    \hspace*{-5mm}
    \epsfxsize=7.0truein
    \epsfbox{mighell.fig04.eps}
    \vspace*{-4cm}
    \caption[]{\baselineskip 1.15em \fig4cap}
  \end{figure}
}
\fi

\def\fig5cap{
\label{fig-ngc361}
The $V$ \vs $\bmv$ color-magnitude diagram of the observed stellar
field in the SMC populous cluster NGC 361.
{\bf{(a)}} The 5172 stars with
signal-to-noise ratios S/N$\geq$10 in
both filters are plotted (dots) along with the
1420 CCD/image defects (open circles).
{\bf{(b)}} The 1607 stars found on the PC1 CCD.
{\bf{(c)}} The 3565 stars found on the WF2, WF3, and WF4 CCDs.
{\bf{(d)}} The ``cleaned'' color-magnitude diagram of NGC 361
contains 1241 stars.
The error bars indicate rms ($1\sigma$)
uncertainties for a single star at the corresponding magnitude.
}

\ifundefined{showfigs}{
  \figcaption[]{\fig5cap}
}\else{
  \clearpage
  \newpage
  \begin{figure}[p]
    \figurenum{5}
    \vspace*{-2cm}
    \hspace*{-5mm}
    \epsfxsize=7.0truein
    \epsfbox{mighell.fig05.eps}
    \vspace*{-4cm}
    \caption[]{\baselineskip 1.15em \fig5cap}
  \end{figure}
}
\fi

\def\fig6cap{
\label{fig-kron3}
The $V$ \vs $\bmv$ color-magnitude diagram of the observed stellar
field in the SMC populous cluster Kron 3.
{\bf{(a)}} The 7669 stars with
signal-to-noise ratios S/N$\geq$10 in
both filters are plotted (dots) along with the
1990 CCD/image defects (open circles).
{\bf{(b)}} The 2596 stars found on the PC1 CCD.
{\bf{(c)}} The 5073 stars found on the WF2, WF3, and WF4 CCDs.
{\bf{(d)}} The ``cleaned'' color-magnitude diagram of Kron 3
contains 2102 stars.
The error bars indicate rms ($1\sigma$)
uncertainties for a single star at the corresponding magnitude.
}

\ifundefined{showfigs}{
  \figcaption[]{\fig6cap}
}\else{
  \clearpage
  \newpage
  \begin{figure}[p]
    \figurenum{6}
    \vspace*{-2cm}
    \hspace*{-5mm}
    \epsfxsize=7.0truein
    \epsfbox{mighell.fig06.eps}
    \vspace*{-4cm}
    \caption[]{\baselineskip 1.15em \fig6cap}
  \end{figure}
}
\fi

\def\fig7cap{
\label{rgb_fits}
Color-magnitude diagrams for 6 of the 7 SMC clusters discussed in this
paper showing the horizontal and red giant branches. The
rectangles indicate the stars used in the calculation of
$V_{\rm{RHB}}$ while the solid lines in each panel show the polynomial fits
resulting from the iterative 2$\sigma$ rejection technique.
}

\ifundefined{showfigs}{
  \figcaption[]{\fig7cap}
}\else{
  \clearpage
  \newpage
  \begin{figure}[p]
    \figurenum{7}
    \vspace*{-70truemm}
    \hspace*{20truemm}
    \epsfxsize=4.5truein
    \epsfbox{mighell.fig07.eps}
    \vspace*{5truemm}
    \caption[]{\baselineskip 1.15em \fig7cap}
  \end{figure}
}
\fi

\def\fig8cap{
\label{fig-slopes}
Metallicity as a function of the slope of the red giant branch.
The primary ({\protect\small\bf{filled circles}})
and secondary ({\protect\small\bf{open circles}})
calibrators from
Sarajedini \& Layden (\protect\cite{sala1997})
are shown.
}

\ifundefined{showfigs}{
  \figcaption[]{\fig8cap}
}\else{
  \clearpage
  \newpage
  \begin{figure}[p]
    \figurenum{8}
    \vspace*{0truemm}
    \hspace*{20truemm}
    \epsfxsize=4.0truein
    \epsfbox{mighell.fig08.eps}
    \vspace*{5truemm}
    \caption[]{\baselineskip 1.15em \fig8cap}
  \end{figure}
}
\fi

\def\fig9cap{
\label{fig-iso}
The V vs $B\!-\!V$ color-magnitude diagrams of the SMC populous
clusters Lindsay 113, Lindsay 1, and NGC 121 (from top to bottom in order
of increasing age and decreasing metallicity).
Bertelli \et (\protect\cite{beet1994})
theoretical isochrones for the indicated ages and metallicities
are shown for comparision.
}

\ifundefined{showfigs}{
  \figcaption[]{\fig9cap}
}\else{
  \clearpage
  \newpage
  \begin{figure}[p]
    \figurenum{9}
    \vspace*{0truemm}
    \hspace*{36truemm}
    \epsfxsize=2.75truein
    \epsfbox{mighell.fig09.eps}
    \vspace*{5truemm}
    \caption[]{\baselineskip 1.15em \fig9cap}
  \end{figure}
}
\fi

\def\fig10cap{
\label{fig-iso2}
The V vs $B\!-\!V$ color-magnitude diagrams of the SMC populous
clusters Kron 3, NGC 339, NGC 416, and NGC 361
(increasing age and decreasing metallicity
from left to right and top to bottom).
Bertelli \et (\protect\cite{beet1994})
theoretical isochrones for the indicated ages and metallicities
are shown for comparision.
}

\ifundefined{showfigs}{
  \figcaption[]{\fig10cap}
}\else{
  \clearpage
  \newpage
  \begin{figure}[p]
    \figurenum{10}
    \vspace*{0truemm}
    \hspace*{26truemm}
    \epsfxsize=4.0truein
    \epsfbox{mighell.fig10.eps}
    \vspace*{5truemm}
    \caption[]{\baselineskip 1.15em \fig10cap}
  \end{figure}
}
\fi

\def\fig11cap{
\label{fig-cmds}
The $M_V$ vs $(\bmv)_o$ color-magnitude diagrams of the intermediate-age
Small Magellanic Cloud populous clusters compared with the fiducial
sequences for Lindsay 1 ($\feh = -1.35$) and Palomar 14 ($\feh = -1.6$).
The clusters are plotted (top to bottom, left to right) in order
of increasing age (see Table \protect\ref{tbl-clpar}).
We have used the Lindsay 1 fiducial derived from the photometry of
Olszewski \et (\protect\cite{olet1987}),
the transformed $BV$ Lindsay 113 photometry of
{\sc{PaperI}}
which was derived from the $BR\,$ photometry
of Mould \et (\protect\cite{moet1984}), and
the Palomar 14 fiducial derived
from the photometry of Sarajedini (\protect\cite{sa1997}).
}

\ifundefined{showfigs}{
  \figcaption[]{\fig11cap}
}\else{
  \clearpage
  \newpage
  \begin{figure}[p]
    \figurenum{11}
    \vspace*{-0mm}
    \hspace*{+10mm}

    \epsfxsize=5.0truein
    \epsfbox{mighell.fig11.eps}
    \vspace*{-0cm}
    \caption[]{\baselineskip 1.15em \fig11cap}
  \end{figure}
}
\fi

\def\fig12cap{
\label{fig-feh_vs_age}
A plot of $\feh$  \vs age for Magellanic Cloud star clusters
with precise metallicity and age determinations.  Our estimates
({\protect\small\bf{filled circles}})
for Lindsay 113, Kron 3, NGC 339, NGC 416, NGC 361,
Lindsay 1, NGC 121 (in order of increasing age)
are combined with other SMC cluster data ({\protect\small\bf{open circles}})
from Da Costa \& Hatzidimitriou (\protect\cite{daha1998}).
The points marked as {\protect\large$\times$} are LMC clusters;
the data for the oldest clusters are from
Olsen \et (\protect\cite{olet1998}) and
the data for clusters younger than 10 Gyr are from
Geisler \et (\protect\cite{geet1997})
and Bica \et (\protect\cite{biet1998}).
The remaining three points
are the present-day abundance of the SMC taken from
Luck \& Lambert (\protect\cite{lula1992}: {\protect\small\bf{asterisk}}),
Russell \& Bessell (\protect\cite{rube1989}: {\protect\small\bf{diamond}}),
and
Hill (\protect\cite{hi1997}: {\protect\small\bf{square}}).
The errorbars reflect our estimates of the uncertainties in $\feh$ and age.
}

\ifundefined{showfigs}{
  \figcaption[]{\fig12cap}
}\else{
  \clearpage
  \newpage
  \begin{figure}[p]
    \figurenum{12}

    \epsfxsize=6.0truein
    \epsfbox{mighell.fig12.eps}
    \vspace*{-0cm}
    \caption[]{\baselineskip 1.15em \fig12cap}
  \end{figure}
}
\fi

\def\fig13cap{
\label{fig-amr_models}
{\bf{Upper panel:}}
The age-abundance data for the 7 clusters discussed
in this paper ({\protect\small\bf{filled circles}}) is compared
with the closed box continuous star-formation model ({\protect\small\bf{solid
curve}})
computed by Da Costa \& Hatzidimitriou (1998, private communication)
for an assumed present day metallicity of
$-0.6$ dex for the SMC.
The younger SMC cluster data ({\protect\small{open circles}}) is from
Da Costa \& Hatzidimitriou (\protect\cite{daha1998}).
The remaining three points
are the present-day abundance of the SMC taken from
Luck \& Lambert (\protect\cite{lula1992}: {\protect\small{asterisk}}),
Russell \& Bessell (\protect\cite{rube1989}: {\protect\small{diamond}}),
and
Hill (\protect\cite{hi1997}: {\protect\small{square}}).
The errorbars reflect our estimates of the uncertainties in $\feh$ and age.
{\bf{Lower panel:}}
This panel is identical with the upper one except that
the SMC bursting model ({\protect\small\bf{solid curve}}) of
Pagel \& Tautvai\u{s}ien\.{e} (\protect\cite{pata1998})
is depicted.
}

\ifundefined{showfigs}{
  \figcaption[]{\fig13cap}
}\else{
  \clearpage
  \newpage
  \begin{figure}[p]
    \figurenum{13}
    \vspace*{-15truemm}
    \hspace*{+15truemm}

    \epsfxsize=4.0truein
    \epsfbox{mighell.fig13.eps}
    \vspace*{+5truemm}
    \caption[]{\baselineskip 1.15em \fig13cap}
  \end{figure}
}
\fi

\def\fig14cap{
\label{fig-clusters_field}
The $V$ \vs $\bmv$ color-magnitude diagrams for the
{\em{HST}} observations of the
SMC populous clusters discussed in this paper.
The total WFPC2 exposure times were $\lea$15 min;
total spacecraft time for these {\em{HST}} observations was $\lea$27 min
per cluster.
The black dots are probable cluster members found on the PC1 CCD
and the gray dots are stars found on the
WF CCDs.
}

\ifundefined{showfigs}{
  \figcaption[]{\fig14cap}
}\else{
  \clearpage
  \newpage
  \begin{figure}[p]
    \figurenum{14}
    \vspace*{-2.5cm}
    \hspace*{+5mm}

    \epsfxsize=8.0truein
    \epsfbox{mighell.fig14.eps}
    \vspace*{-4cm}
    \caption[]{\baselineskip 1.15em \fig14cap}
  \end{figure}
}
\fi

\end{document}